\documentclass[review]{article}
\usepackage{thesis}
\usepackage{lineno,hyperref}
\usepackage[font={singlespacing,small,it}]{caption}

\begin{document}

%\begin{frontmatter}

\title{Green neighbourhoods in low voltage networks: measuring impact of electric vehicles and photovoltaics on load profiles}%\tnoteref{mytitlenote}}
%\tnotetext[mytitlenote]{This work is funded by TVV project.}
\date{}
\newcommand{\address}{Centre for the Mathematics of Human Behaviour, Department of Mathematics and Statistics, University of Reading, UK}
\author{Laura Hattam\footnote{l.hattam@reading.ac.uk, \address},\quad Danica Vukadinovi\'c Greetham\footnote{d.v.greetham@reading.ac.uk, \address} }
%\author[mysecondaryaddress]{Stephen Haben}
%\ead{s.haben@maths.ox.ac.uk}

%\address[mysecondaryaddress]{Mathematical Institute, University of Oxford, Woodstock Road, Oxford OX2 6GG, UK}

\maketitle

\begin{abstract}
In the near future various types of low-carbon technologies (LCTs) are expected to be widely employed throughout the United Kingdom. However, the effect that these technologies will have at a household level on the existing low voltage (LV) network is still an area of extensive research. We propose an agent based model that estimates the growth of LCTs within local neighbourhoods, where social influence is imposed. Real-life data from a LV network is used that comprises of many socially diverse neighbourhoods. Both electric vehicle uptake and the combined scenario of electric vehicle and photovoltaic adoption are investigated with this data. A probabilistic approach is outlined, which determines lower and upper bounds for the model response at every neighbourhood. This technique is used to assess the implications of modifying model assumptions and introducing new model features. Moreover, we discuss how the calculation of these bounds can inform future network planning decisions.
\end{abstract}

%\newcommand{\keywords}{
%low voltage networks modelling\sep electric vehicles\sep photovoltaics \sep Monte Carlo simulations}
%\keywords

%\end{frontmatter}

\section{Introduction}

From $2000$ onwards there is a tendency of national electric demand  (so called baseload) in UK and other developed countries to stagnate or even to decrease despite the population increase. The UK energy statistics  show total electricity consumption year on year decreasing on UK, GB and south-east level \footnote{\url{https://www.gov.uk/government/uploads/system/uploads/attachment_data/file/296183/pn_march_14.pdf}} and  governmental statistics tables \footnote{\url{https://www.gov.uk/government/statistics/energy-consumption-in-the-uk}(see Domestic data tables – Table 3.07) } present a consistent UK trend from 2008-2014 showing a fall in a domestic consumption before and after temperature adjustment. This is mostly due to  different regulations (requiring more energy efficient appliances, phasing out incandescent light bulbs in EU and other countries, etc.) that gave rise to technology innovations in energy efficiency, LED lighting, loft and cavity insulations etc. The current predictions are that the UK domestic baseload will continue to decrease in the next ten to fifteen years \footnote{\url{http://www2.nationalgrid.com/UK/Industry-information/Future-of-Energy/Electricity-Ten-Year-Statement/}} due to better efficiency of electric appliances and lighting \cite{Bilton:2014,Bossman:2015}. The anticipated increase in the domestic demand will come mostly from the new builds and low carbon technologies employed in existing buildings.

Electrification of transport and heating forecasted for the near future in order to reduce carbon footprint are predicted to be the main contributors  to increased electric demand. The whole picture is made more complex by distributed generation and variability of renewable sources of energy  that result in new peaks and troughs in aggregated consumption. Not all projected changes threaten to worsen this situation. A big mitigating factor is energy storage that can help smoothen generation and demand and offer cost efficient local solutions.

We are interested in measuring the combined impact of electric vehicles and solar panels on low voltage networks.
Several possible issues with low voltage networks that might arise from the described smart grid developments are already recognised: frequent peak loads that reduce headway, voltage drops, phase imbalance, etc. Due to the complexity of human societies, any predictions on uptake of electric vehicles (EVs)  and photovoltaics (PVs) comes with large uncertainties. Different models of uptake exist and are used by network planners, but by their nature it is quite difficult to validate such models, and to decouple influence of different decisions when modelling.

Our contribution is two-fold. Firstly, we present an agent based model of load profiles for the uptake of LCTs in local neighbourhoods when social influence is present. Our neighbourhoods are based on real-life low voltage networks containing multiple substations and feeders. This model uses a sample of realistic EV and PV profiles to simulate future uptake. Secondly, we demonstrate techniques that allow for a thorough mathematical analysis of results. Probabilistic methods based on multiple simulations allow the calculation of upper and lower bounds for the model response, which we refer to as confidence bounds. These bounds are used to understand the inner-workings of the model and to measure the effects of introducing/changing the model's parameters.

The outline of the paper is the following. In Section \ref{prev} we give an overview of the recent relevant results. In Section \ref{mod} the model is described in detail, and the data that we use for initialisation and calibration of the model. The focus is initially EV adoption only. In Section \ref{cbs} the confidence bounds creation is explained with some simulation results shown. Then in Section \ref{ctbs} the adaptation of the model to include socio-demographic information is discussed. As well, confidence bounds are used to assess the impact of this new model feature. Next in Section \ref{evpv} the model is further modified to investigate the combined uptake of EVs and PVs. Again, confidence bounds are computed to determine the effect of changing our model assumptions. Finally in Section \ref{Concl} we discuss the implications of our results and their possible use in design, planning and policy.
\section{Previous Work}\label{prev}
There is fast growing literature (see reviews \cite{Yong:2015}, \cite{Elnozahy:2013}) on different impacts of EVs and PVs on future power grids with regard to load profile, system losses, voltage profile, phase unbalance, harmonic and stability impact. We focus on the impact on load profile and on low voltage (LV) networks.

\subsection{Impact of PVs and EVs to LV networks}
For EVs, most of the existing work is based on predictions, simulations or small pilot trials data. Only recently larger data sets based on trials are becoming available.
Focusing mostly on LV networks impact, in \cite{Neaimeh:2015} the authors create generic local networks to assess the impact of EVs charging on neighbourhoods.

In \cite{Watson:2016}, the authors aimed to measure the impact of PVs on LV networks in New Zealand. They were looking in particular at over-voltage and overload of conductors and transformers. They created a power-flow model of an LV network and simulated varying percentages of PV uptake. PV was based on a specific installation  with an output power of 3.7kW. These uniform PVs were then randomly distributed through different parts of the LV network classified as rural, urban, industrial and city. Their results show that only very high PV penetration (over 45\%) will cause an overload of conductors and in most cases overvoltage is not much higher than the existing statutory limit. In a review \cite{Elnozahy:2013}, the major technical impacts of small PV installations are discussed: excessive reverse power flow, overvoltages along distribution feeders, increased difficulty of voltage control, increased power losses (caused by reverse power flow), severe phase unbalance, and so on.

A microgrid case-study from a neighbourhood in Utrecht in Netherlands, looking at the combination of PVs and EVs throughout a year is described in \cite{VanderKam:2015}. Based on simulations, but using February demand projected over the whole year,  the authors compared several control algorithms. Their results show a potential for relative peak reduction and increased self-consumption when using smart charging and vehicles to grid technology.

In \cite{Navarro:2016} Monte-Carlo simulations are used to measure the impact of several low carbon technologies including EVs and PVs. Similarly to our approach, the authors use a realistic LV network with $7$ feeders and sample from realistic profiles for baseload and for LCTs. Although the network area examined here is much larger with $44$ feeders considered. Their focus is on identifying thermal and voltage problems in different feeders. While they use random allocation of LCT, we compare random allocation with a clustered one using some socio-demographic information, but we focus on load only.
\subsection{Agent based modelling of PVs and EVs uptake}
In \cite{Poghosyan:2015} a simple agent-based model of EV uptake and their impact on local grid is given, based on governmental scenarios of future EV uptake for UK and a small pilot project where participants were incentivised to charge over night.  As expected, having a variety of EV charging patterns helped to reduce the peaks as opposed when all the domestic charging happened after work or over night. Comparing a random and clustered uptake simulations have shown that some local grids might see a substantial increase of peak loads faster than expected.

In \cite{Dehoog:2014} EVs charging was simulated on two real low-voltage networks (semi-rural and urban) to show that  spatial distribution of loads is an important consideration when analysing voltage drops and network stability.

In \cite{Adepetu:2016} an agent-based model using San Francisco as a test city is presented considering how different policies and battery technologies might affect the uptake and usage of EVs.  The model includes a set of agents with socio-demographic properties  and attitudes and  EV ecosystem with costs of gas, electricity, rebates and public charging stations. Each three months, agents consider if they need a new vehicle. Based on their properties, attitudes and state of their social network they acquire (or not) an EV to use for their daily commute. The  social network is created randomly based on similarities in age, income and residential locations  of agents. This allows for exploring different scenarios (for example, increasing or decreasing rebate for EVs, installing charging at work stations, increasing battery sizes, etc) and looking at the impacts on average daily load.

An interesting review of the research on adoption of EVs is given in \cite{Rezvani:2015}. The authors warn about a tendency to generalise results obtained from surveys of people who do not possess EVs, and that the intention-behaviour gap (projecting intentions to adopt EV to the actual adoption) should be better understood than it is currently. They also point out that although social influence and `green neighbourhood' effect is already confirmed by different studies, the ways in which green neighborhoods have been formed is not studied.
\section{The agent based model\label{mod}}
Here, the impact of future LCT adoption is predicted with our agent based model that applies a clustered distribution of technologies to a sample UK population (Bracknell, UK). The clustering follows the Joneses effect, one of the causes for `green neighbourhoods' where a household is more likely to acquire a LCT if their neighbour has one. We model EVs and PVs as they are visible outside.

\begin{figure}
\begin{center}
\includegraphics[width=2.5in]{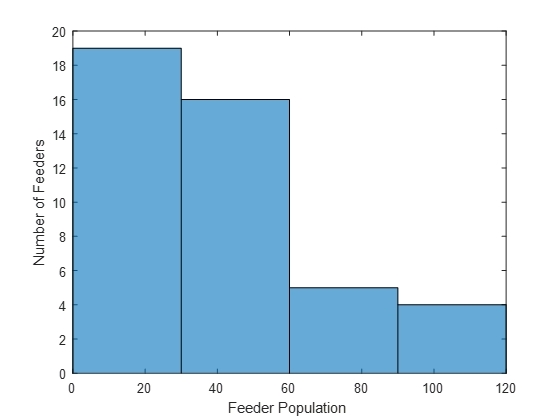}
\caption{The distribution of feeder populations (the number of households along a feeder), where there are $44$ feeders and $1848$ properties in total.
\label{feedpop}}
\end{center}
\end{figure}

Our network is based on a realistic LV network and comprises of $44$ feeders. The household population at each feeder varies, where the minimum number of households along one feeder is $4$ and the maximum is $114$. Figure \ref{feedpop} demonstrates the variation in feeder size. Note that each feeder corresponds to one neighbourhood and all households along a particular feeder are considered neighbours. Overall, there are $1848$ properties, where $7$ are households with PVs installed and $71$ are commercial properties.

We also have three data sets (baseloads) that were created using metered data from this LV network. This information was collected on Thursday the $15^{\text{th}}$ of January $2015$ (winter), Thursday the $7^{\text{th}}$ of May $2015$ (spring) and Thursday the $9^{\text{th}}$ of July $2015$ (summer). The data sets consist of a combination of metered and predicted daily demand energy profiles (kWh) for every household, where a genetic algorithm was used to allocate monitored endpoints to unmonitored customers (see \cite{gia16}). These profiles have readings every half hour and therefore for each household we have a profile load as a vector of length $48$.

\begin{figure}
\begin{center}
\includegraphics[width=4.8in]{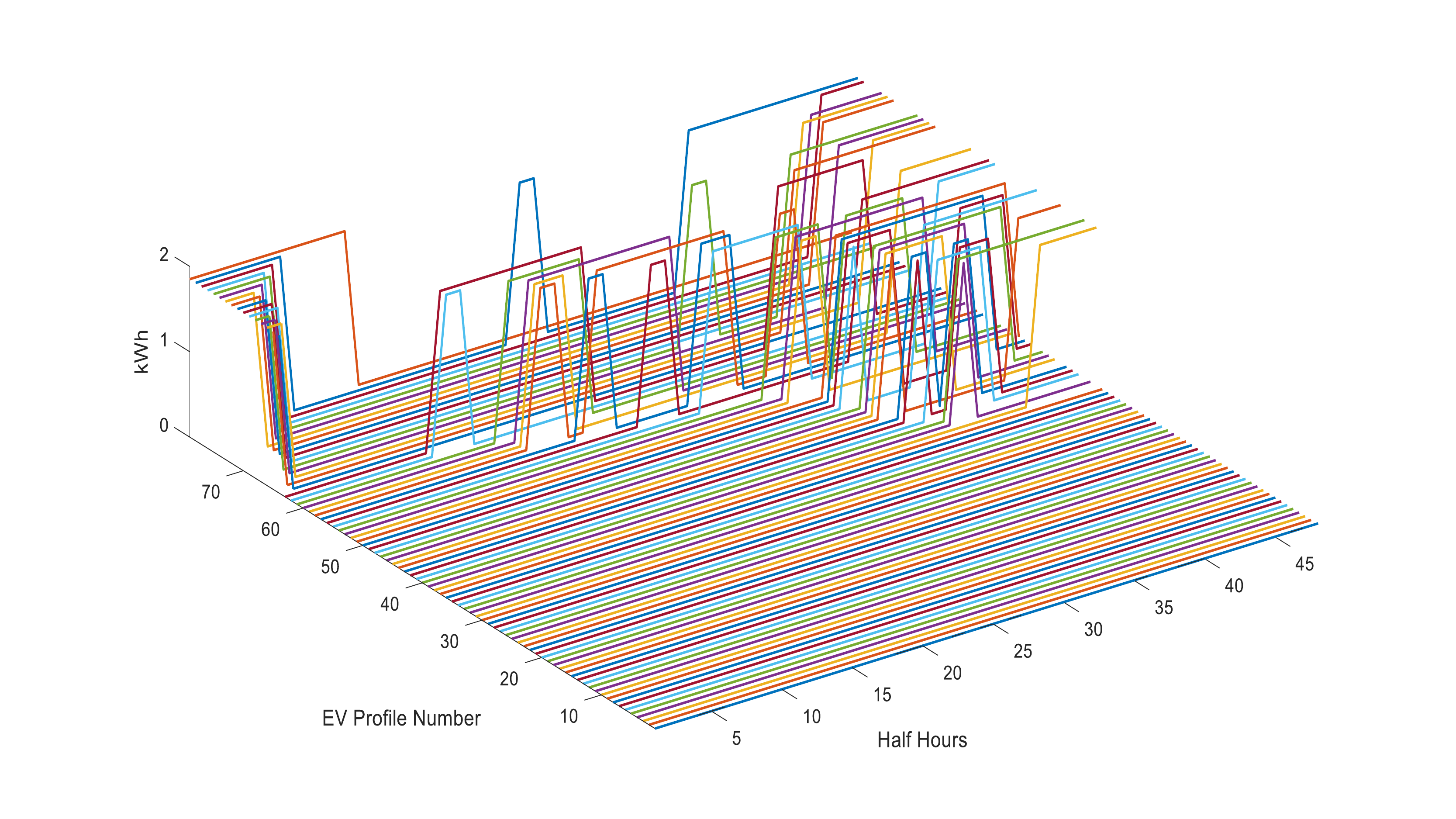}
\includegraphics[width=4.8in]{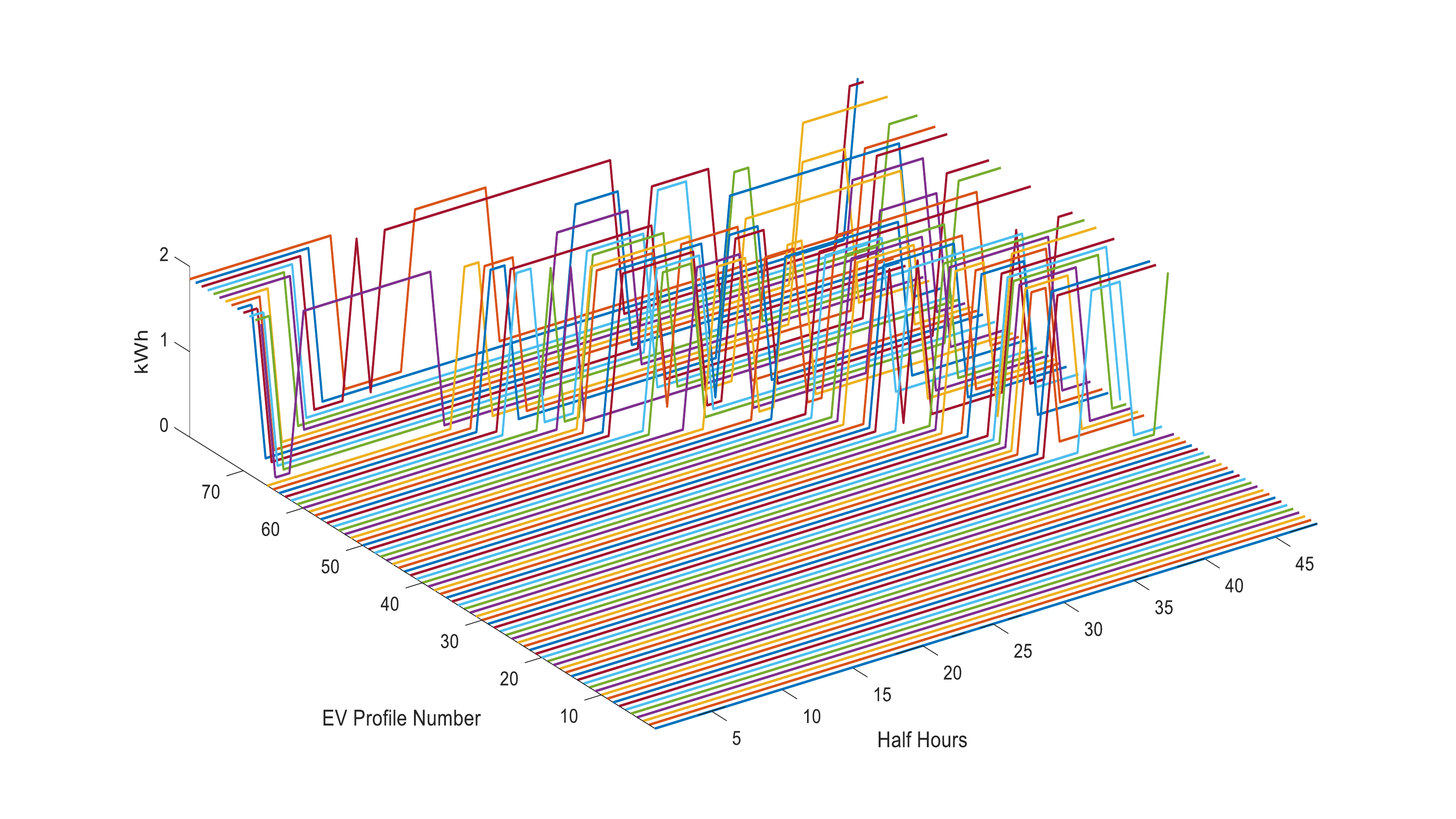}\\
\includegraphics[width=4.8in]{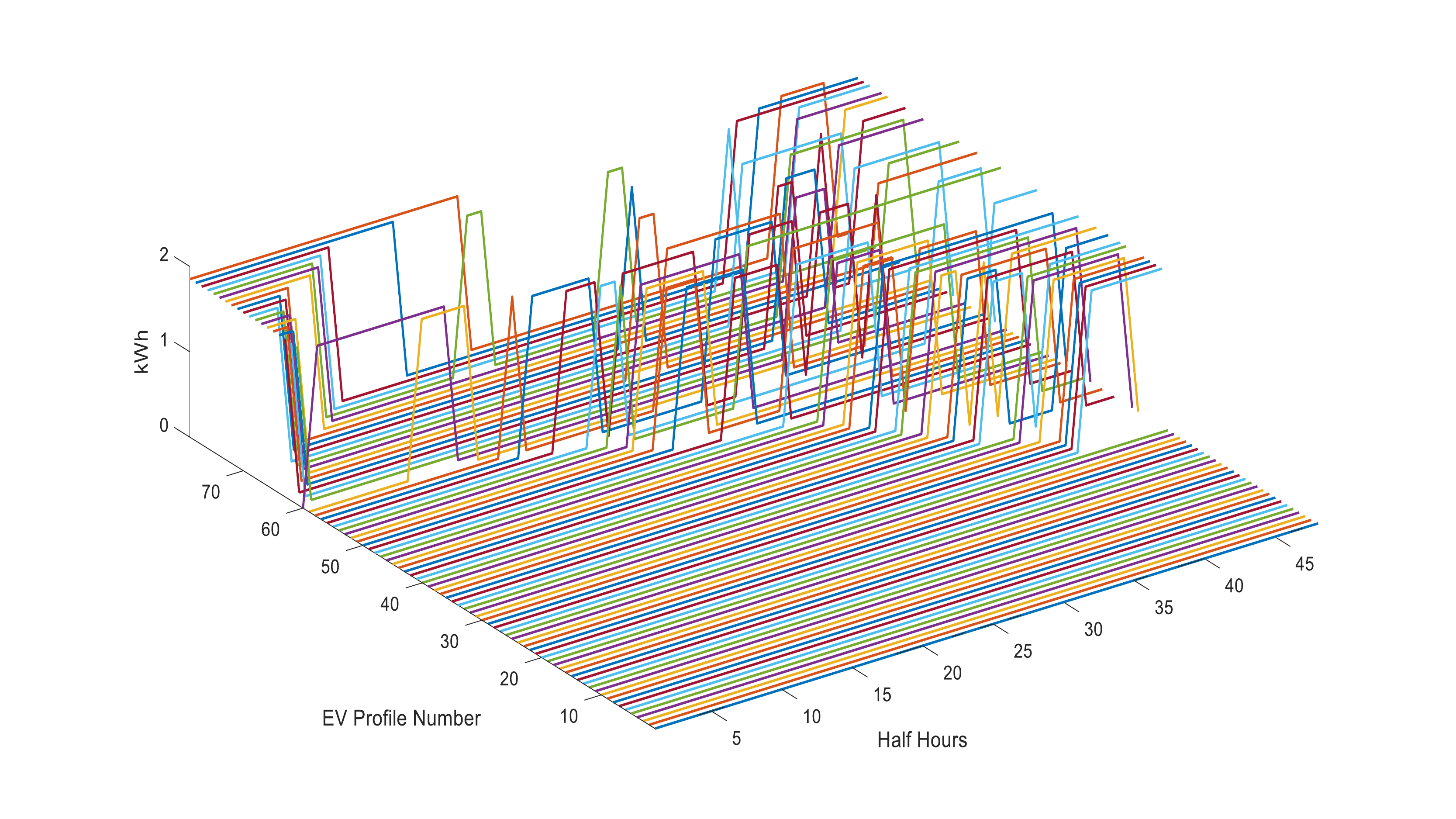}
\caption{The EV daily profiles used for modelling EV uptake. These profiles correspond to three days during the My Electric Avenue trial, which are $08/05/2014$ (top), $10/07/2014$ (middle) and $15/01/2015$ (bottom).
\label{evprofile}}
\end{center}
\end{figure}

Initially, the focus here is the clustered allocation of EVs, although later the combination of EV and PV uptake is investigated. The EV charging profiles used in our model were generated during the $55$ week trial conducted by My Electric Avenue \footnote{The data is available from \url{http://www.eatechnology.com/products-and-services/myelectricavenue}}, where the number of participants increased as the trial progressed. These profiles consist of the two values `$0$kWh' and `$1.85$kWh', which represent when the EV is not charging and charging respectively. They have readings also every half hour. Three days from this trial are selected, which are Thursday May the $8^{\text{th}}$ 2014 (week $16$ of the trial), Thursday July the $10^{\text{th}}$ $2014$ (week $25$ of the trial) and Thursday January the $15^{\text{th}}$ $2015$ (week $52$ of the trial). These dates are chosen since they correspond seasonally to the baseload dates. There are $79$ households that consistently participate during weeks $16-52$ of the trial and therefore, we have $79$ EV daily profiles that are representative of winter, spring and summer charging behaviour. The $79$ profiles for each season are shown in Figure \ref{evprofile}. The zero profiles suggest that these households did not charge their EV on this particular day.

The following outlines the clustering algorithm applied to forecast EV uptake:
\begin{itemize}
\item{Firstly, we establish the percentage of households in the sample population that will adopt EVs and the number of years it will take (here, this is set to $8$ years to simulate a typical distribution network operator's planning period).}
\item{Next, an initial random distribution of EV seeds is performed to simulate the first year of EV uptake.}
\item{Then, during the remaining years, EVs are assigned to households according to the score $s$ (refer to (\ref{scoredef})).}
\item{The number of EV households (households that adopted an EV) increases linearly every year until the specified amount is attained.}
\item{Lastly, EV profiles are assigned to the EV households, where
$\textit{EV household profile}=\textit{baseload}+\textit{EV profile}.$
}
\end{itemize}
It is important to note that all $71$ commercial properties in our data set never receive LCTs as our focus is household uptake. As well, there is one feeder comprised of only commercial properties, so this site is given zero LCT load always.

\begin{figure}
\begin{center}
\includegraphics[width=2.8in]{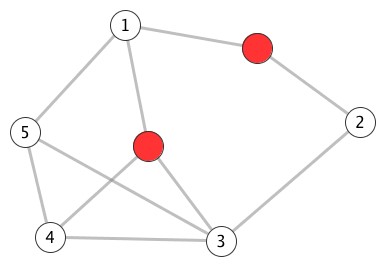}
\includegraphics[width=2.8in]{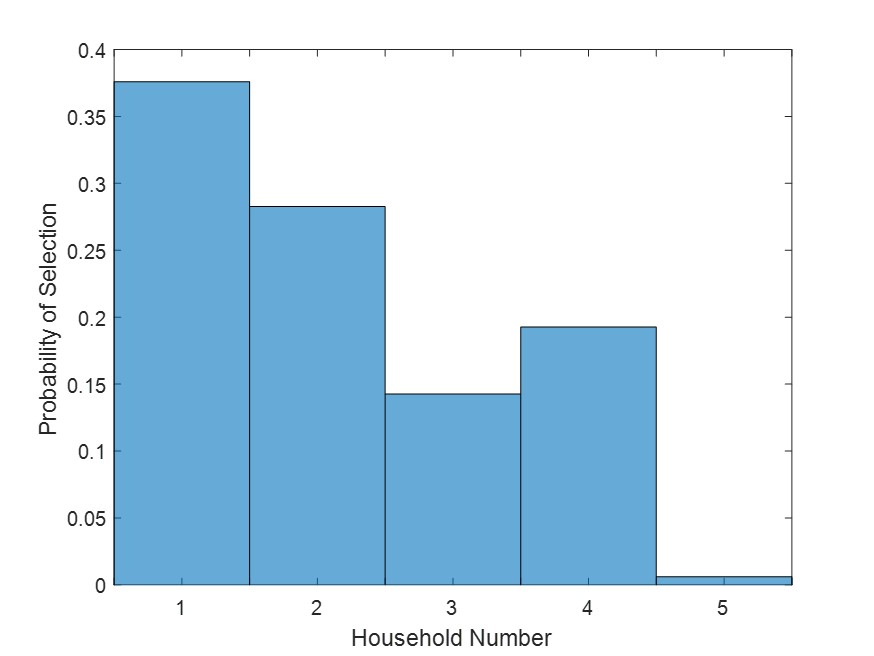}
\caption{Left: Depiction of a simplified, small network. The red circles signify EV allocation and the numbered circles correspond to eligible households. The connecting lines indicate neighbours. Right: Their probability of selection for EV assignment by the random number generator - Household 1 has the largest probability of EV assignment as two of its three neighbours already have EVs. \label{prob_s}}
\end{center}
\end{figure}

The score, $s$, assigned to eligible households is the percentage of PVs and EVs in its neighbourhood presently, where
\beq
s=100\times\left(\frac{\text{Number of neighbours with an EV and/or PV}}{\text{Number of neighbours}+1}\right).\label{scoredef}
\eeq
This score is proportional to the probability of selection by a random number generator. Figure \ref{prob_s} illustrates the selection process with a simplified network, which comprises of red and numbered circles that represent EV and eligible households respectively. Also shown is their probability of EV assignment by the random number generator. Once a household is selected, they become an EV household for the remaining years of the simulation, with $s$ updated every year. Using $s$ to inform EV allocation leads to clusters of EVs forming around the initial seeds. This method is an adaptation of the algorithm proposed in \cite{Poghosyan:2015}, which was also applied to model EV uptake.

There is an assumed link between increased neighbourhood diversity and a heavily populated feeder. As a result, when transforming these larger sites into greener neighbourhoods, the impact from one EV household should be comparatively small. To account for this, $s$ depends upon the feeder population and therefore, the influence of one household on its neighbours is relative to the neighbourhood size.

Note that when a household is given an EV, the EV profile assigned to them is randomly selected from $79$ possible profiles. If the baseload applied is representative of spring, summer or winter then the EV profile chosen will also correspond to spring, summer or winter respectively.

\section{Confidence bounds\label{cbs}}
For some fixed model parameters, there are many feasible outcomes. This is due to the initial random distribution of seeds highlighting different neighbourhoods every model run. As well, one of $79$ possible EV profiles are randomly assigned to households, causing further variation in the model result. Consequently, we aim to determine upper and lower bounds of the model response for a fixed set of parameters, which we label confidence bounds. These bounds will be calculated by undertaking $500$ consecutive model runs and will therefore relate to the EV load variance, not the baseload. Since the clustering is based on neighbourhoods, which are defined by feeders, the bounds will be computed at each feeder.

The following details the method used to calculate confidence bounds:
\begin{itemize}
\item{Specify the model parameters, which are the uptake percentage and the number of years i.e. $30\%$ EV uptake ensures $\lceil 0.3\times1848\rceil$ properties receive an EV each simulation.}
\item{Complete $500$ simulations.}
\item{After each simulation, record the aggregate result at the feeder. The $44$ feeders are considered together so that $0-100\%$ of households along a particular feeder can receive an EV each simulation.}
\item{The aggregate data is then used to calculate $10\%,\;50\%$ and $90\%$ quantiles at the feeder. The feeder lower and upper bounds correspond to the $10\%$ and $90\%$ quantiles.}
\item{The quantile with the baseload subtracted represents the variation in EV load at the feeder. Then, dividing the quantiles by the number of households along the feeder, we can compare the $44$ feeders and their EV loads.}
\end{itemize}

\begin{figure}
\begin{center}
\includegraphics[width=2.13in]{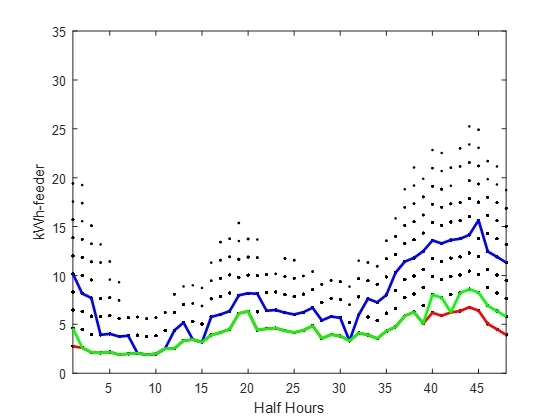}
\includegraphics[width=2.13in]{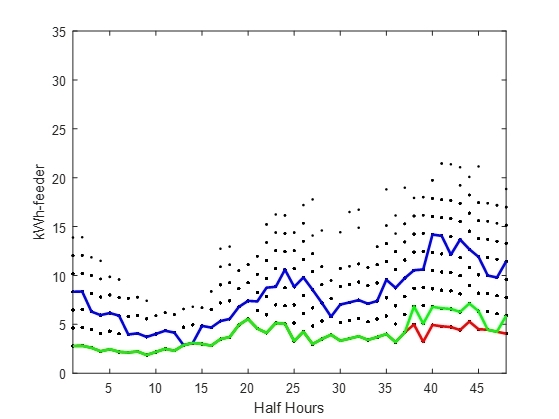}
\includegraphics[width=2.13in]{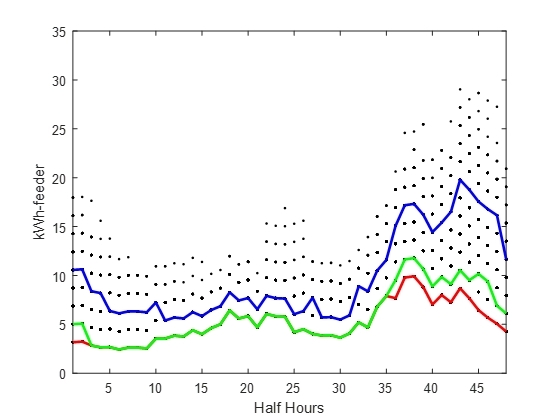}\\
\includegraphics[width=2.13in]{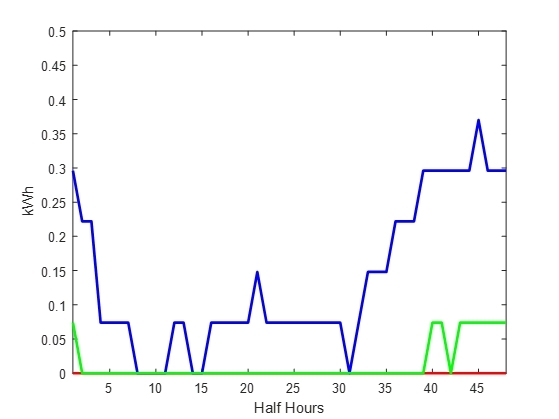}
\includegraphics[width=2.13in]{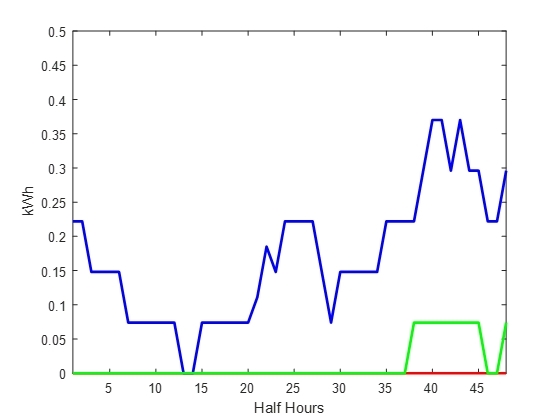}
\includegraphics[width=2.13in]{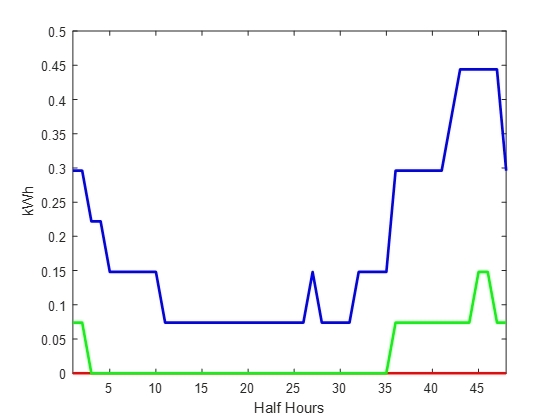}\\
\includegraphics[width=2.13in]{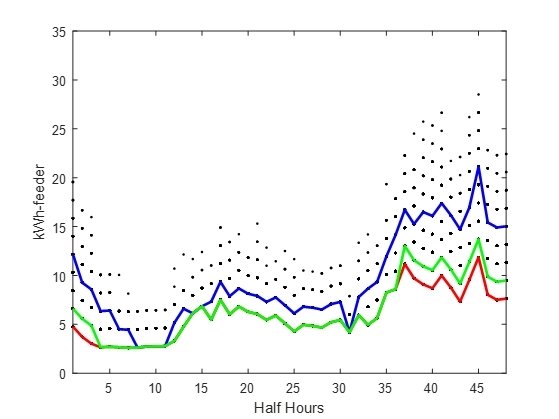}
\includegraphics[width=2.13in]{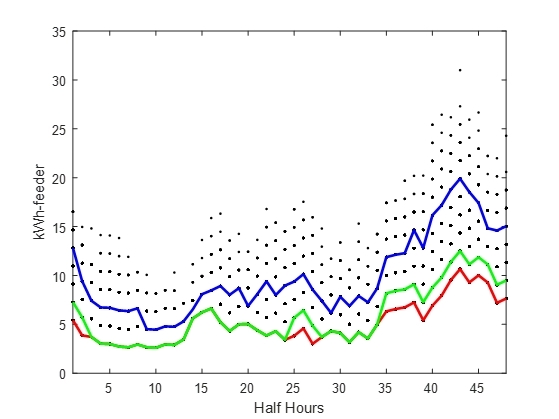}
\includegraphics[width=2.13in]{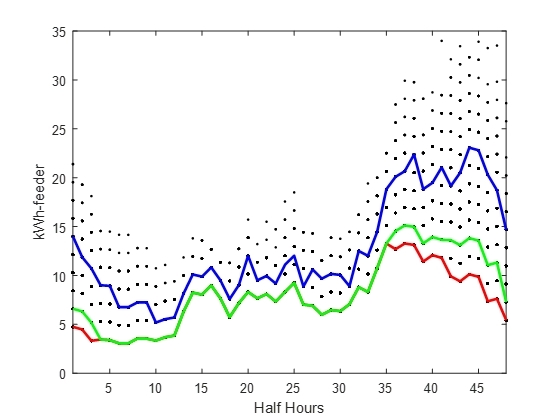}\\
\includegraphics[width=2.13in]{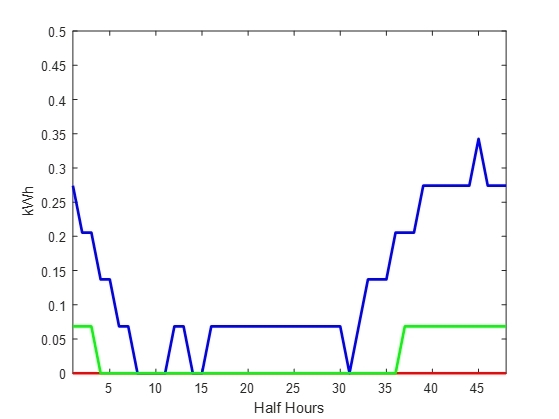}
\includegraphics[width=2.13in]{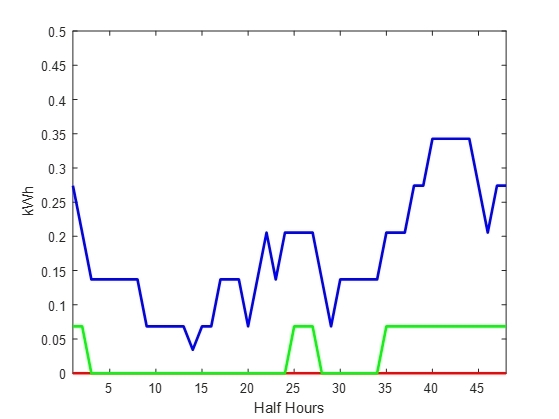}
\includegraphics[width=2.13in]{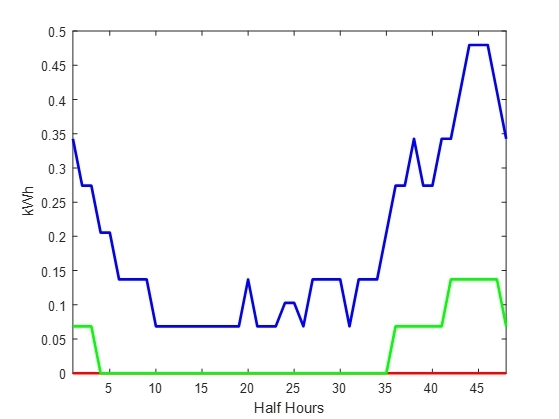}
\caption{The result of $500$ simulations with $30\%$ EV uptake (random seed allocation). These are the confidence bounds for feeders $15$ and $17$, where the red, green and blue curves represent the $10\%$, $50\%$ and $90\%$ quantiles respectively. First and second row: Feeder $15$ has $25$ households, $0$ PV properties, $1$ commercial property. Third and fourth row: Feeder $17$ has $27$ households, $0$ PV properties, $0$ commercial properties. Left: Spring. Middle: Summer. Right: Winter. First and third row: The aggregate result at the feeder. Second and fourth row: The EV quantile (the baseload subtracted), divided by the number of houses along the feeder.\label{f31cb}}
\end{center}
\end{figure}

\begin{figure}
\begin{center}
\includegraphics[width=2.13in]{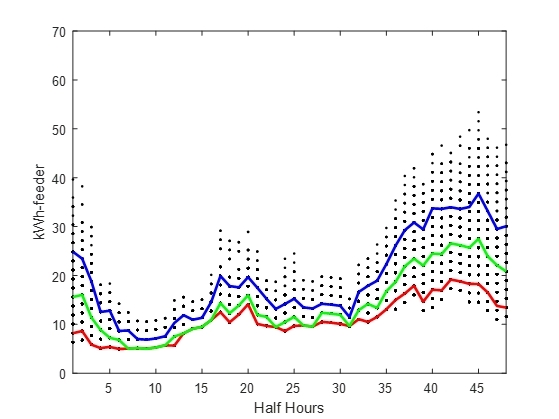}
\includegraphics[width=2.13in]{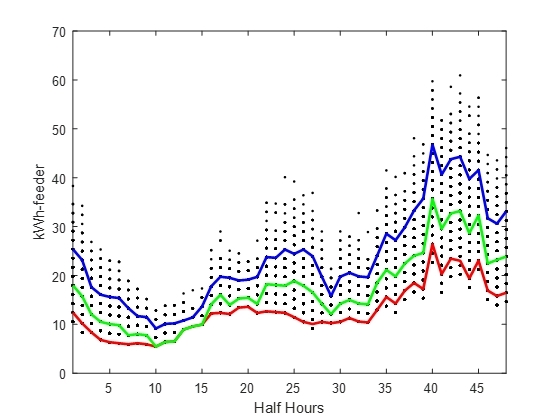}
\includegraphics[width=2.13in]{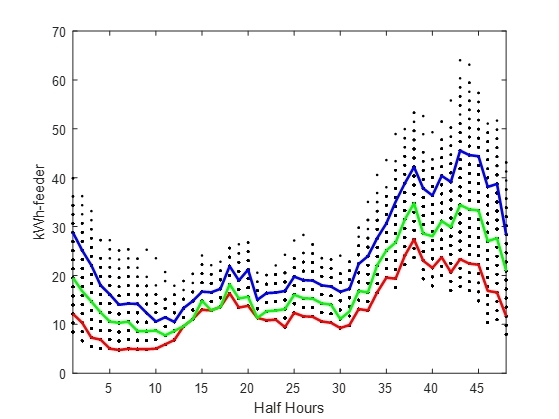}\\
\includegraphics[width=2.13in]{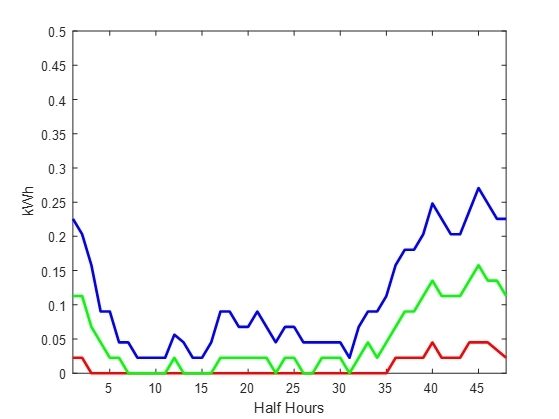}
\includegraphics[width=2.13in]{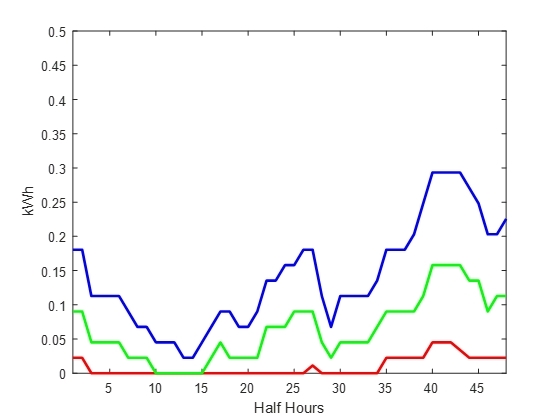}
\includegraphics[width=2.13in]{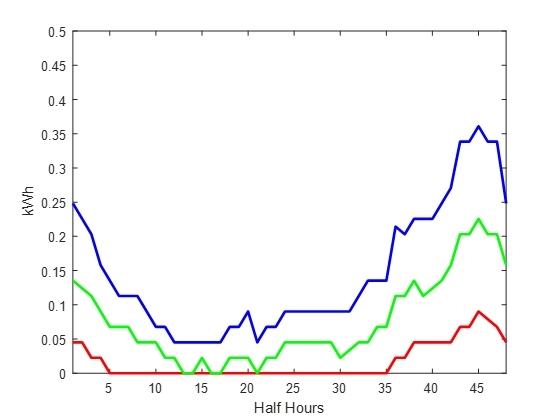}\\
\includegraphics[width=2.13in]{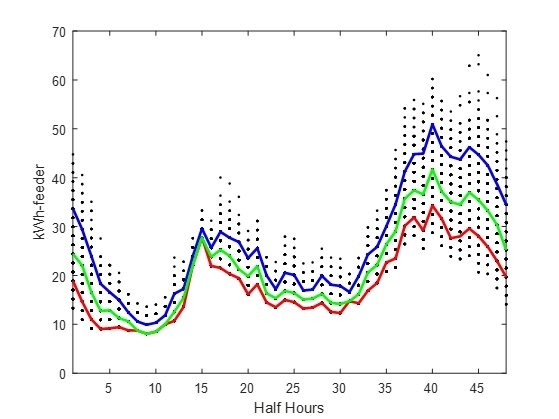}
\includegraphics[width=2.13in]{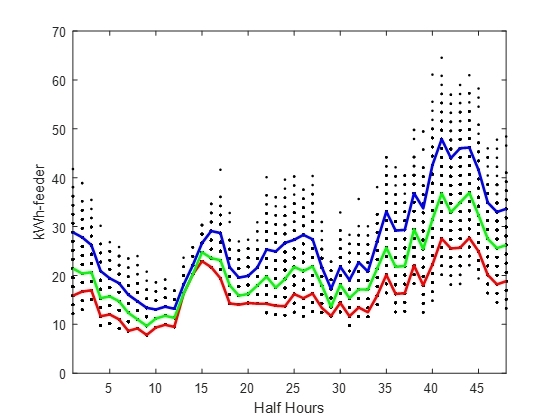}
\includegraphics[width=2.13in]{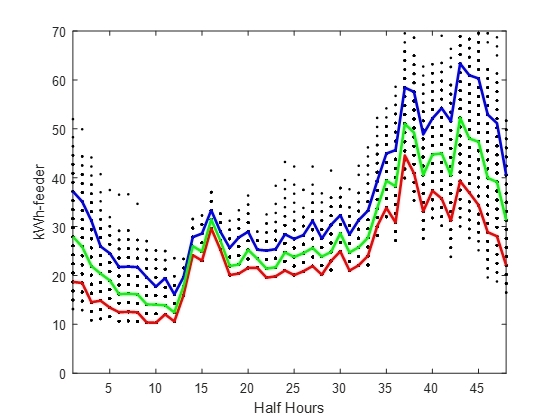}\\
\includegraphics[width=2.13in]{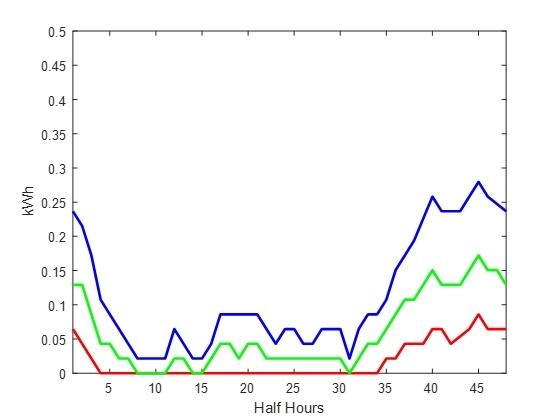}
\includegraphics[width=2.13in]{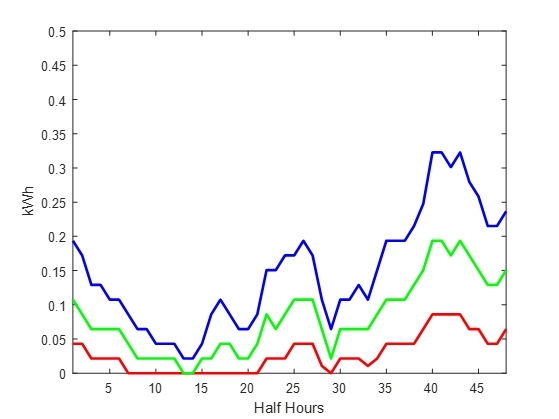}
\includegraphics[width=2.13in]{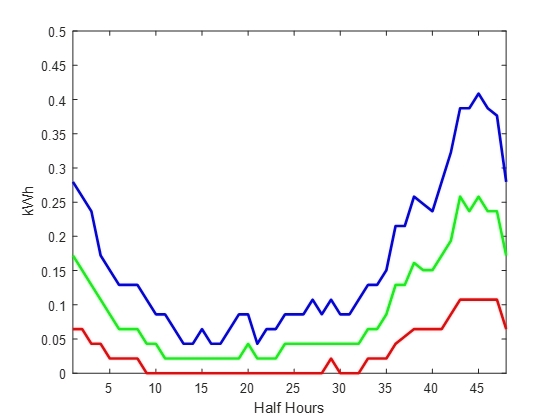}
\caption{The result of $500$ simulations with $30\%$ EV uptake (random seed allocation). These are the confidence bounds for feeders $39$ and $40$, where the red, green and blue curves represent the $10\%$, $50\%$ and $90\%$ quantiles respectively. First and second row: Feeder $39$ has $82$ households, $0$ PV properties, $1$ commercial property. Third and fourth row: Feeder $40$ has $86$ households, $1$ PV property, $2$ commercial properties. Left: Spring. Middle: Summer. Right: Winter. First and third row: The aggregate result at the feeder. Second and fourth row: The EV quantile (the baseload subtracted), divided by the number of houses along the feeder.\label{f4cb}}
\end{center}
\end{figure}

In Figures \ref{f31cb} and \ref{f4cb}, the results for feeders $15$, $17$, $39$ and $40$ are depicted. The top and third panels display the aggregate result at the feeder (including baseload), where the black dots represent the response from $500$ model runs. The red, green and blue curves are the $10\%,\;50\%$ and $90\%$ quantiles respectively calculated from the black dots. The second and bottom panels show the quantiles with the baseload subtracted, divided by the number of households along the feeder. The left, middle and right figures correspond to spring, summer and winter respectively, where, for example, the spring result uses the spring baseload and spring EV profiles discussed in Section \ref{mod}. Comparing the second and bottom panels of Figure \ref{f31cb}, as well as the second and bottom panels of Figure \ref{f4cb}, it is apparent that feeders with similar household numbers receive comparable EV loads. Furthermore, less populated feeders have greater EV peaks, demonstrated by the blue curves. This can be attributed to increased neighbourhood diversity when the feeder population is larger and therefore, it is more difficult to influence your neighbours and to form an EV majority. Also, from analysing the quantiles in Figures \ref{f31cb} and \ref{f4cb}, it is evident that the spread between the $10\%$ and $50\%$ trends is far less notable for smaller feeders. Moreover, the red curve sits along the baseload in Figure \ref{f31cb}. This suggests that less populated feeders do often avoid EV assignment. The weather also influences the EV result, since there is an obvious variation in EV charging behaviour across the three seasons. In particular, comparing summer to winter using the upper bound, we see that there is increased activity during summer days and larger peaks during winter nights.


\section{Adding socio-demographic information\label{ctbs}}
The simulations performed in Section \ref{cbs} randomly allocated EV seeds. Next, we aim to improve our model by introducing council tax band (CTB) information to instead inform seed distribution. Here, it is assumed that larger homes correspond to higher CTBs. We note here that other socio-demographic information  can be used if available (for example Acorn \footnote{\url{http://acorn.caci.co.uk/}}). We use CTB as it is publicly available \footnote{\url{https://www.gov.uk/council-tax-bands}} and it allows us to identify neighbourhoods that have a higher percentage of larger properties.

A survey of Californian EV owners \cite{ev13} revealed that generally they owned and lived in single family dwellings that had parking and space to install a charging point. They also had higher incomes, which typically relates to living in larger homes. Furthermore, present EV owners commonly had a PV installed at their property. Also acknowledged was that neighbour influence was an important factor in EV adoption since clusters of EV households had formed in California. This study therefore supports initialising the seeds guided by CTB information and then imposing neighbour influence to determine the growth of EV ownership.

Hence, our model is now adapted to firstly favour PV properties and larger households, signified by higher CTBs. To implement this, we assign to every household an initial score, $s_i$, such that
\beq
s_i=\frac{100 \operatorname{ctb}_{hh}^j}{8^j},\label{sidef}\eeq
where $ctb_{hh}=1,2,\ldots,8$ when the household's CTB is $A,B,\ldots,H$ respectively and $j$ is some positive integer. As well, PV (resp. commercial) properties are given the score $s_i=100$ (resp. $s_i=0$). The score is proportional to the likelihood of selection by a random number generator. This relationship is consistent with that portrayed in Figure \ref{prob_s}. It should be noted that $s_i$ is only used during the first year when seeds are allocated, then $s$ as given by (\ref{scoredef}),  applies for the remaining years.

Choosing $j$ determines how dependent seed assignment is on the CTB information, where CTB influence increases with $j$. Here we set $j=4$.

\begin{figure}
\begin{center}
\includegraphics[width=2.5in]{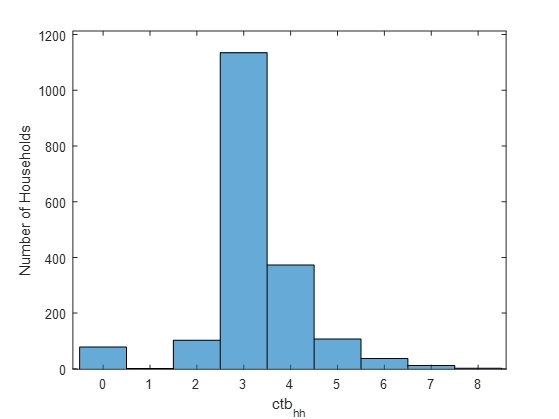}
\includegraphics[width=2.5in]{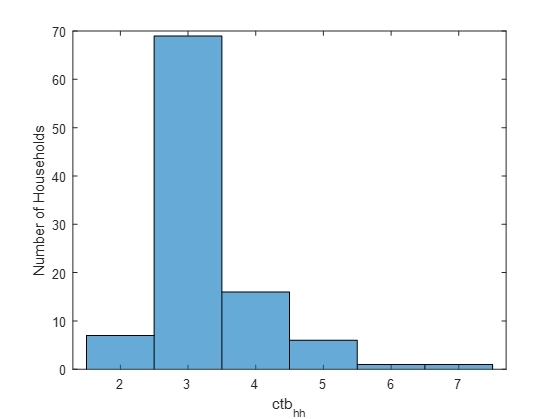}\\
\includegraphics[width=2.5in]{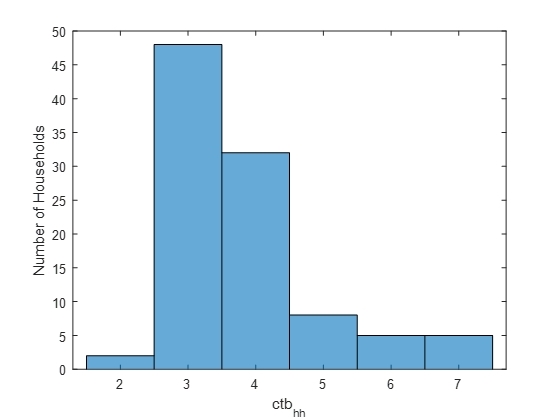}
\includegraphics[width=2.5in]{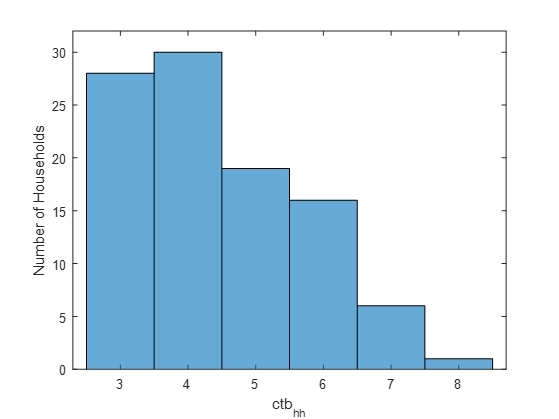}
\caption{Histograms demonstrating the spread of household council tax bands, where $ctb_{hh}=1,2,\ldots,8$ represents household council tax band $A,B,\ldots,H$ respectively (commercial properties are given $ctb_{hh}=0$). Top left: CTBs of the $1848$ properties. Top right: CTBs of $100$ random seeds. Bottom left: CTBs of $100$ seeds distributed using (\ref{sidef}) with $j=2$. Bottom right: CTBs of $100$ seeds distributed using (\ref{sidef}) with $j=4$.\label{ctbhist}}
\end{center}
\end{figure}

In Figure \ref{ctbhist}, a histogram depicts the spread of household CTBs within our sample population. As well, a comparison is given of $100$ seeds randomly selected (top right) and those guided by CTB, with $j=2$ (bottom left) and $j=4$ (bottom right). These figures display the $ctb_{hh}$ of the 100 nominated households. We propose that by setting $j=4$, the subsequent initial EV population reflects the survey findings \cite{ev13}. This is due to over $70\%$ of seeds now having a CTB greater than $C$, where the grouping $A-C$ typically represents smaller dwellings. Although, $j$ is a model parameter that can be varied.

\begin{figure}
\begin{center}
\includegraphics[width=2.13in]{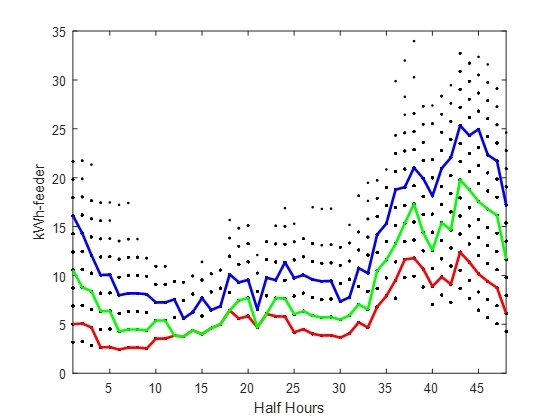}
\includegraphics[width=2.13in]{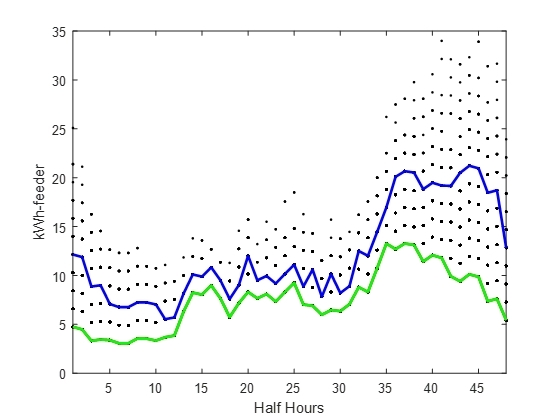}\\
\includegraphics[width=2.13in]{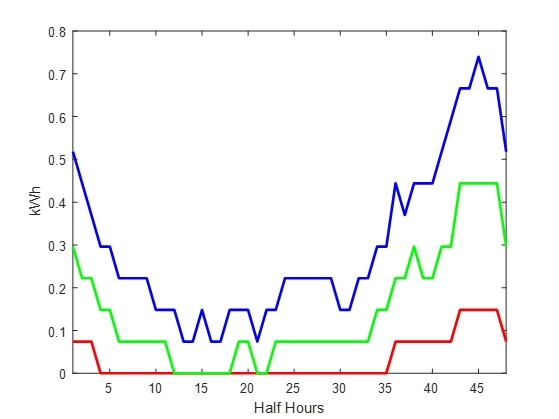}
\includegraphics[width=2.13in]{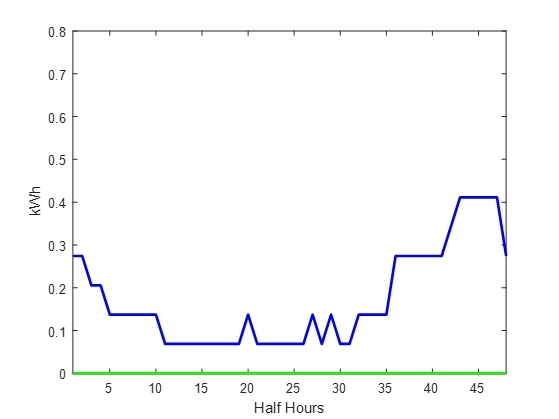}\\
\includegraphics[width=2.13in]{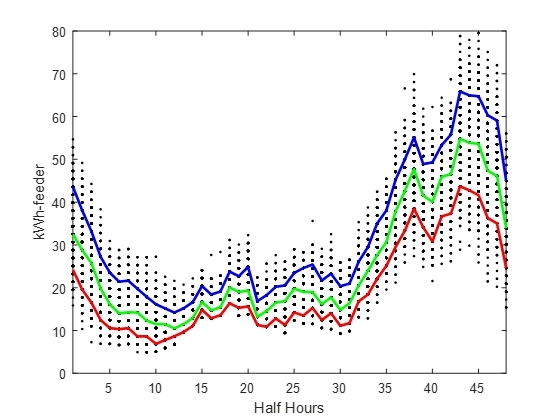}
\includegraphics[width=2.13in]{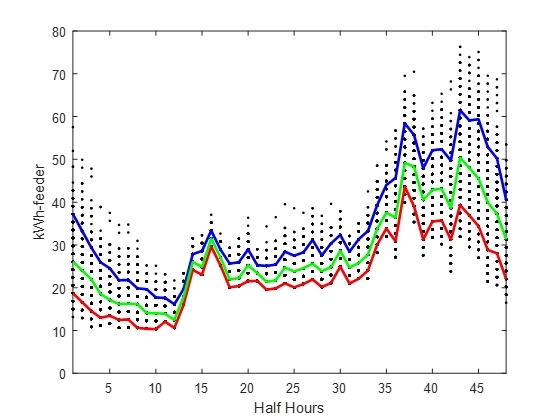}\\
\includegraphics[width=2.13in]{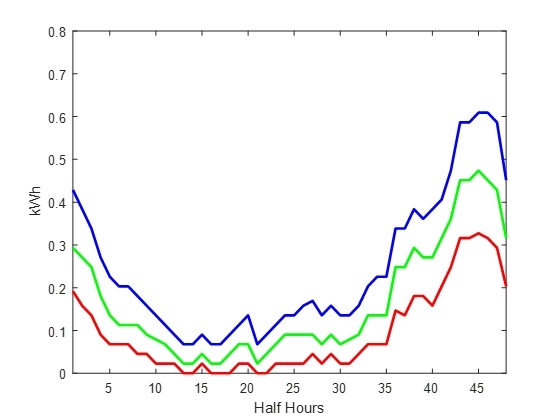}
\includegraphics[width=2.13in]{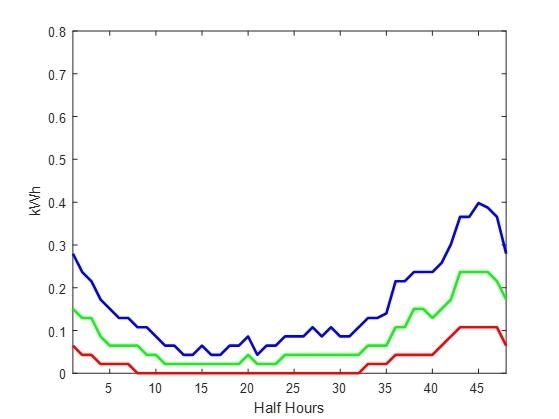}
\caption{The result of $500$ simulations with $30\%$ EV uptake winter (seed allocation informed by CTB). These are the confidence bounds for select feeders, where the red, green and blue curves represent the $10\%$, $50\%$ and $90\%$ quantiles respectively. First and second row, left: Feeder $15$ has $25$ households, $0$ PV properties, $1$ commercial property and the number of households with CTB$>D=60\%$ and CTB$>C=100\%$. First and second row, right: Feeder $17$ has $27$ households, $0$ PV properties, $0$ commercial properties and the number of households with CTB$>D=0\%$ and CTB$>C=19\%$. Third and fourth row, left: Feeder $39$ has $82$ households, $0$ PV properties, $1$ commercial property and the number of households with CTB$>D=54\%$ and CTB$>C=93\%$. Third and fourth row, right: Feeder $40$ has $86$ households, $1$ PV property, $2$ commercial properties and the number of households with CTB$>D=0\%$ and CTB$>C=9\%$. First and third row: The aggregate result at the feeder. Second and fourth row: The EV quantile (the baseload subtracted), divided by the number of houses along the feeder.\label{f1631ctb}}
\end{center}
\end{figure}

Confidence bounds can be used to measure the effect of changing our model assumptions. Here, we analyse the influence of using CTB to inform the initial seed distribution, instead of random initial distribution. The winter results for feeders $15$, $17$, $39$ and $40$ are shown in Figure \ref{f1631ctb}. It is evident that feeders with about the same sized populations are no longer given a similar EV load. The upper bounds depicted along the second panel of Figure \ref{f1631ctb} reveal that feeder $15$ receives a significantly larger load than feeder $17$. This is due to $60\%$ of properties along feeder $15$ having a CTB greater than $D$, whereas for feeder $17$ it is $0\%$. Similarly, the fourth panel of Figure \ref{f1631ctb} suggests that feeder $39$ has been assigned a greater EV load compared to feeder $40$, which is due to $54\%$ of households along feeder $39$ having a CTB of more than $D$, when feeder $40$ has $0\%$. Furthermore, of these four feeders, feeder $15$ overall has the largest EV peak, which is a result of both feeder size and its households' CTBs.

Now that socio-demographic data has been incorporated, higher potential peaks are exhibited at certain feeders than previously predicted. This modelling suggests that these feeders are likely sites for future network issues. Refer to the Appendix for an overview of the results at all $44$ feeders.

\section{The combination of electric vehicles and photovoltaics\label{evpv}}
The additional impact of PV adoption on the electricity network is now examined by adapting our model to also consider PVs.

\begin{figure}
\begin{center}
\includegraphics[width=2.8in]{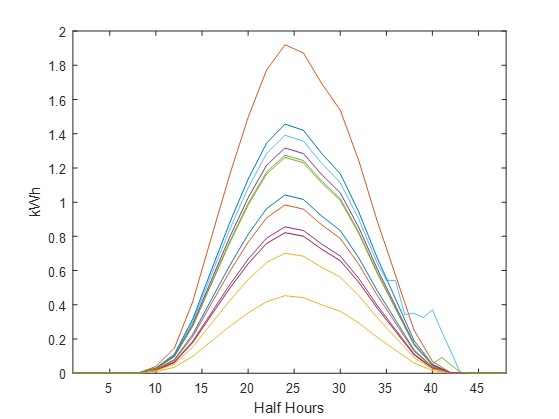}
\caption{The $12$ summer PV generation profiles created using solar radiation data. \label{PV_s}}
\end{center}
\end{figure}

As part of the New Thames Valley Vision project, surplus generation and solar radiation data was recorded at $12$ households with PVs installed. PV daily generation profiles were then created, defined every half hour, by assuming that PV generation is proportional to solar radiation. It should be noted that other influences, such as ambient and surface temperatures, would also contribute to PV generation, but we ignore these to simplify our model. Thus, for our analysis only solar radiation is used. As a result, we obtained three sets of $12$ PV daily generation profiles that were representative of spring, summer and winter generation. For this investigation, only the summer baseload and summer PV profiles are applied to simulate uptake. In Figure \ref{PV_s}, the $12$ summer PV daily generation profiles are shown. These profiles are scaled so that the maximum generation is $1.9$ kWh to comply with UK standards. When a household is given a PV, one of the $12$ possible profiles are randomly selected and then subtracted from their baseload i.e. a household with an EV and a PV is assigned the profile, $\textit{EV+PV household profile}=\textit{baseload}+\textit{EV profile}-\textit{PV profile}$.

The confidence bounds discussed in Section \ref{cbs} are now used to quantify the effect of both EV and PV adoption by our sample population. Simulations for $30\%$ EV and $30\%$ PV uptake are conducted. Firstly, the clustering algorithm outlined in Section \ref{mod} is applied, where the initial seed is randomly distributed. Then, the seed allocation is guided by CTB. There are now two scores assigned to eligible households. These are $s_{EV}$ and $s_{PV}$, where both are defined using (\ref{scoredef}) and updated every year. A household's likelihood for EV (resp. PV) selection by a random number generator is proportional to $s_{EV}$ (resp. $s_{PV}$). The dependence of selection on these scores is demonstrated by Figure \ref{prob_s}.

Again the $71$ commercial properties within our data set do not receive a LCT. Also, we ensure that the $7$ households with PVs installed already are not allocated an additional PV.

\begin{figure}
\begin{center}
\includegraphics[width=2.13in]{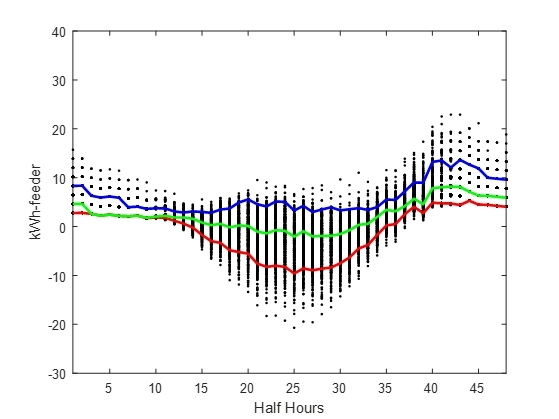}
\includegraphics[width=2.13in]{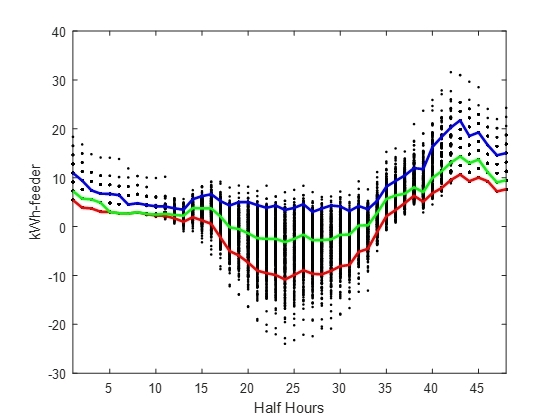}\\
\includegraphics[width=2.13in]{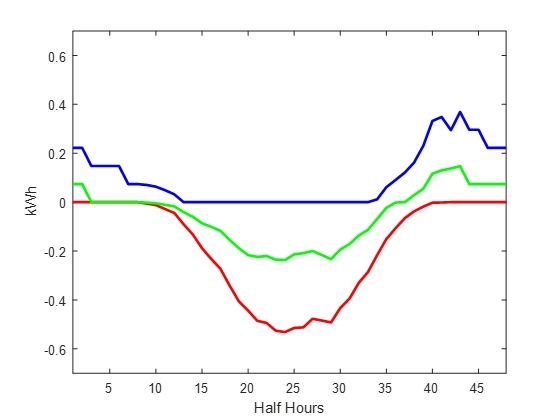}
\includegraphics[width=2.13in]{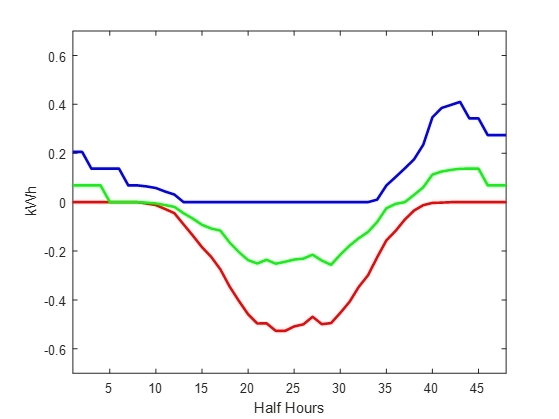}\\
\includegraphics[width=2.13in]{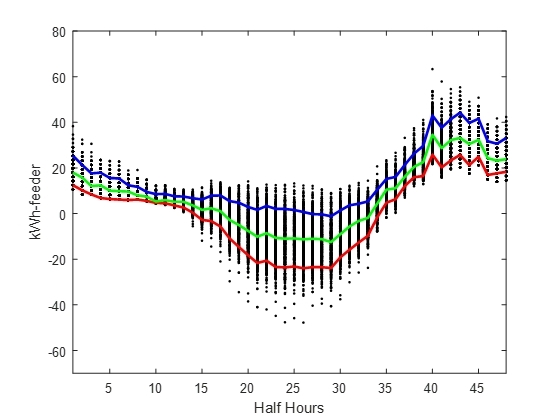}
\includegraphics[width=2.13in]{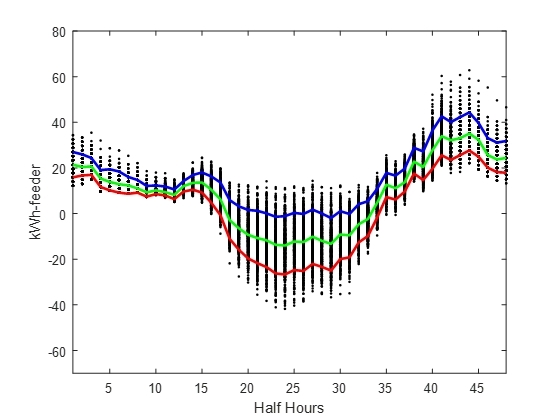}\\
\includegraphics[width=2.13in]{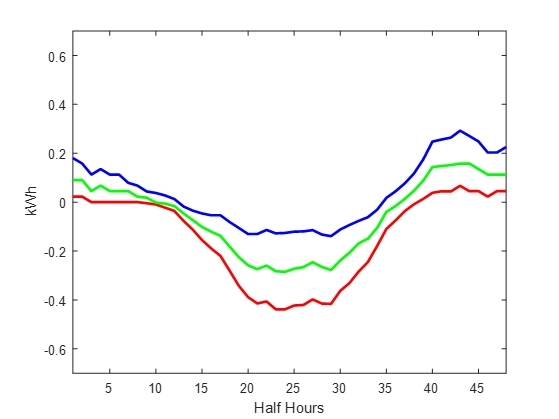}
\includegraphics[width=2.13in]{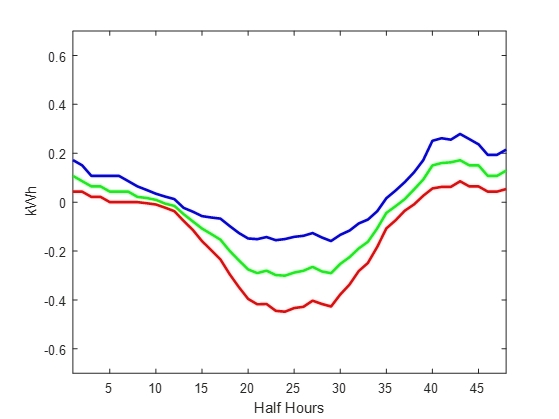}
\caption{The result of $500$ simulations with $30\%$ EV uptake and $30\%$ PV uptake summer (random seed allocation). These are the confidence bounds for select feeders, where the red, green and blue curves represent the $10\%$, $50\%$ and $90\%$ quantiles respectively. First and second row, left: Feeder $15$ has $25$ households, $0$ PV properties, $1$ commercial property and the number of households with CTB$>D=60\%$ and CTB$>C=100\%$. First and second row, right: Feeder $17$ has $27$ households, $0$ PV properties, $0$ commercial properties and the number of households with CTB$>D=0\%$ and CTB$>C=19\%$. Third and fourth row, left: Feeder $39$ has $82$ households, $0$ PV properties, $1$ commercial property and the number of households with CTB$>D=54\%$ and CTB$>C=93\%$. Third and fourth row, right: Feeder $40$ has $86$ households, $1$ PV property, $2$ commercial properties and the number of households with CTB$>D=0\%$ and CTB$>C=9\%$. First and third row: The aggregate result at the feeder. Second and fourth row: The EV/PV quantile (the baseload subtracted), divided by the number of houses along the feeder.\label{EVPV1631}}
\end{center}
\end{figure}

In Figure \ref{EVPV1631}, the results for feeders $15$, $17$, $39$ and $40$ are displayed. Here, the initial seeds have been randomly allocated, where EV and PV seeds are distributed separately. The quantiles depicted along the second and fourth panels reveal significant troughs (red curve) develop during the day and large peaks (blue curve) form at night. Furthermore, it is evident that feeders of a similar size again receive comparable EV/PV loads, where the red trough and blue peak are more prominent for smaller feeders, suggesting increased variability at these feeders.

\begin{figure}
\begin{center}
\includegraphics[width=2.13in]{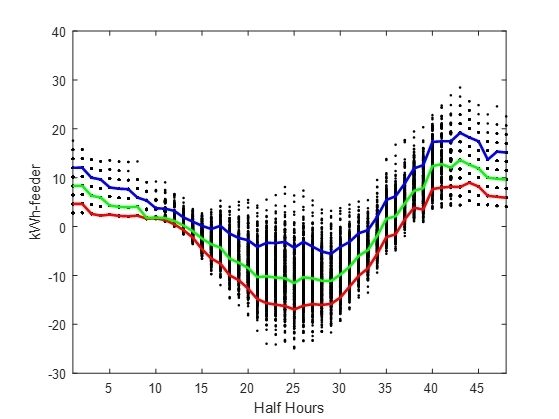}
\includegraphics[width=2.13in]{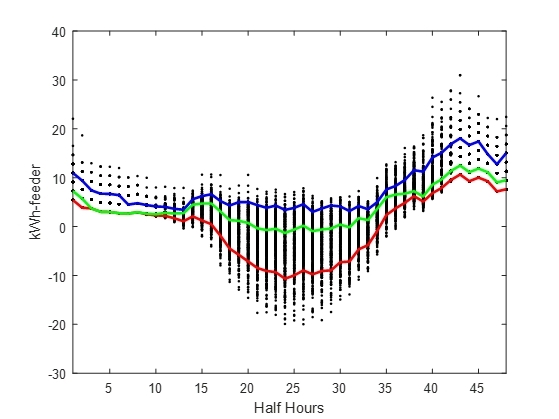}\\
\includegraphics[width=2.13in]{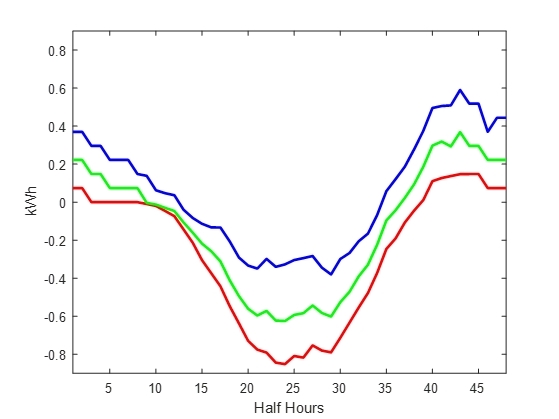}
\includegraphics[width=2.13in]{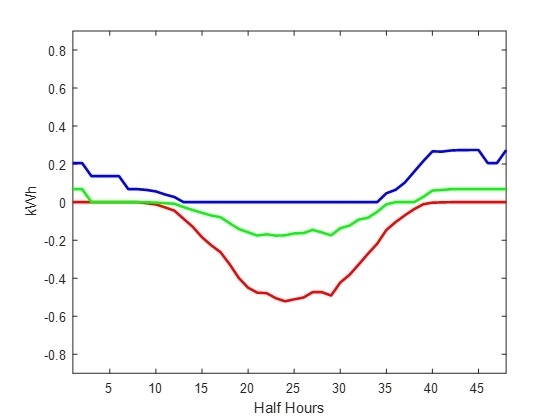}\\
\includegraphics[width=2.13in]{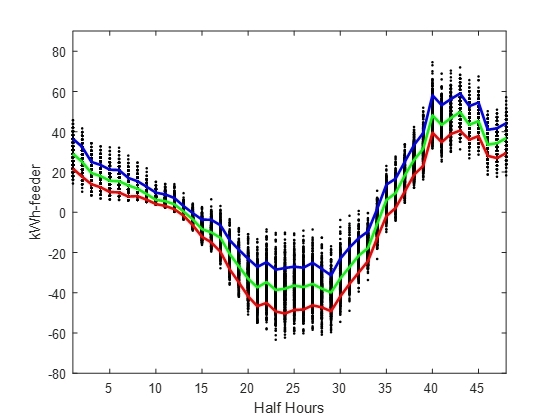}
\includegraphics[width=2.13in]{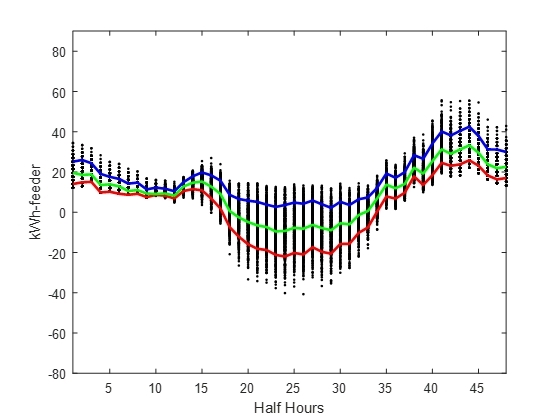}\\
\includegraphics[width=2.13in]{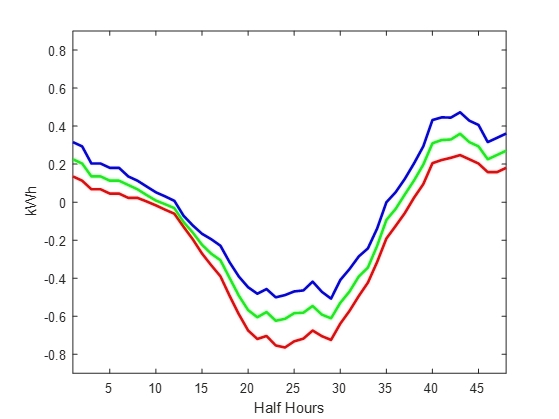}
\includegraphics[width=2.13in]{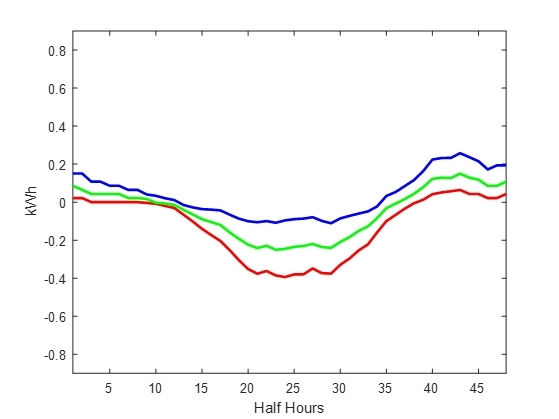}
\caption{The result of $500$ simulations with $30\%$ EV uptake and $30\%$ PV uptake summer (seed allocation informed by CTB). These are the confidence bounds for select feeders, where the red, green and blue curves represent the $10\%$, $50\%$ and $90\%$ quantiles respectively. First and second row, left: Feeder $15$ has $25$ households, $0$ PV properties, $1$ commercial property and households with CTB$>4=60\%$ and CTB$>C=100\%$. First and second row, right: Feeder $17$ has $27$ households, $0$ PV properties, $0$ commercial properties and the number of households with CTB$>D=0\%$ and CTB$>C=19\%$. Third and fourth row, left: Feeder $39$ has $82$ households, $0$ PV properties, $1$ commercial property and the number of households with CTB$>D=54\%$ and CTB$>C=93\%$. Third and fourth row, right: Feeder $40$ has $86$ households, $1$ PV property, $2$ commercial properties and the number of households with CTB$>D=0\%$ and CTB$>C=9\%$. First and third row: The aggregate result at the feeder. Second and fourth row: The EV/PV quantile (the baseload subtracted), divided by the number of houses along the feeder.\label{EVPV1631ctb}}
\end{center}
\end{figure}

Next, in Figure \ref{EVPV1631ctb}, the results for feeders $15$, $17$, $39$ and $40$ are given, where now the seed distribution is informed by CTB information. The allocation of EV and PV seeds are again separate. Consistent with previous findings, due to introducing CTB, feeders $15$ and $39$ have greater extremes. Interestingly, these values are roughly the same for feeders $15$ and $39$, and hence, independent of feeder size. This is a result of now modelling two technologies, which amplifies the clustering effect. Although, the result variability is more pronounced for the smaller feeders, indicated by the spread of the quantiles. As a result of using CTBs, the minimum and maximum loads obtained at feeders $15$ and $39$ are now larger than initially estimated (see Figure \ref{EVPV1631}).

\begin{figure}
\begin{center}
\includegraphics[width=2.13in]{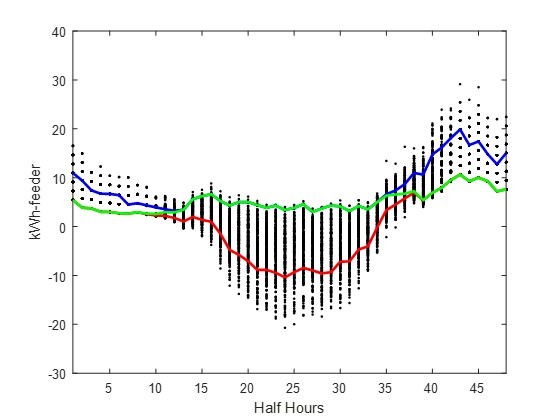}
\includegraphics[width=2.13in]{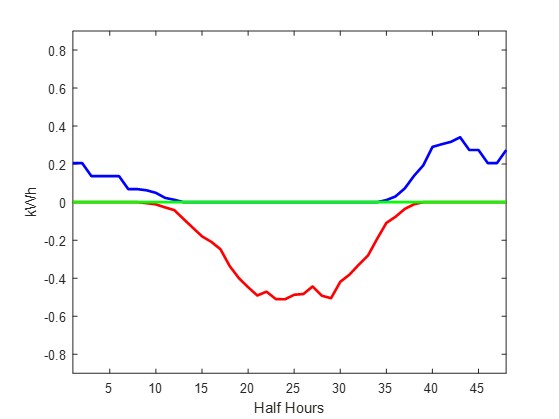}
\caption{The result of $500$ simulations with $30\%$ EV uptake and $30\%$ PV uptake summer (seed allocation informed by CTB). Every household that obtained an EV were also given a PV. These are the confidence bounds for feeders $17$, where the red, green and blue curves represent the $10\%$, $50\%$ and $90\%$ quantiles respectively. Feeder $17$ has $27$ households, $0$ PV properties, $0$ commercial properties and the number of households with CTB$>D=0\%$ and CTB$>C=19\%$. Left: The aggregate result at the feeder. Right: The EV/PV quantile (the baseload subtracted), divided by the number of houses along the feeder.\label{EVPV16last}}
\end{center}
\end{figure}

Lastly, simulations of $30\%$ EV and $30\%$ PV uptake with CTB information are again performed, except now we ensure that all households which receive an EV with our clustering algorithm are also given a PV. The results are actually extremely similar to those depicted in Figure \ref{EVPV1631ctb}. The most significant difference observed is at feeder $17$ and is shown in Figure \ref{EVPV16last}. This is expected as our clustering method already promotes the growth of EV$+$PV groupings.

Modelling EV and PV uptake reveals nighttime peaks and daytime troughs, which enlarge at some feeders when the CTB information is applied. Consequently, these feeders appear as highly likely locations for future network problems. Refer to the Appendix where an overview of the results at the $44$ feeders is given.

\section{Conclusions}\label{Concl}
An agent based model was outlined that considered social factors to predict the uptake of low-carbon technologies. The data used was taken from real-life, with real substation and feeder assignment. This allowed us to sort the $1848$ properties into $44$ realistic neighbourhoods. Then neighbour influence was imposed to determine uptake. The model also applied sets of EV and PV profiles that were representative of spring, summer and winter usage. To assess the model response, a probabilistic approach was proposed that provided feeder confidence bounds. These were a result of $500$ consecutive simulations and therefore, the bounds measured the variation in LCT load. Next, another aspect of social influence was introduced with socio-demographic information also guiding LCT selection. More specifically, we ensured that bigger households were more likely to acquire a LCT. Confidence bounds were then utilised to quantify the effect of implementing this change. In particular, the potential peaks/troughs at select feeders were amplified as these neighbourhoods comprised of clusters of larger homes. The modelling undertaken focussed on EV adoption and then the combination of EV and PV uptake. To investigate these different scenarios and their possible model outcomes, computing confidence bounds was extremely effective. Moreover, the upper bound can also be used to determine the available headroom at each feeder for some specified uptake percentage. Identifying headroom is essential for network planning since negative headroom indicates transmission is greater than the maximum available power, causing issues for the electricity provider. Hence, subsequent to the upper bound calculation, certain feeders can be highlighted as likely sites for network malfunction when subjected to LCT demand. Furthermore, when analysing PV uptake as well, the lower bound becomes an equally important measure since negative power at the feeder level is also problematic for the electricity provider. Confidence bounds therefore will be an important tool to inform new policies and planning so that the future impact of LCTs on the LV network can be minimised.
\section*{Appendix}
\begin{figure}
\begin{center}
\includegraphics[width=6.0in,height=2.5in]{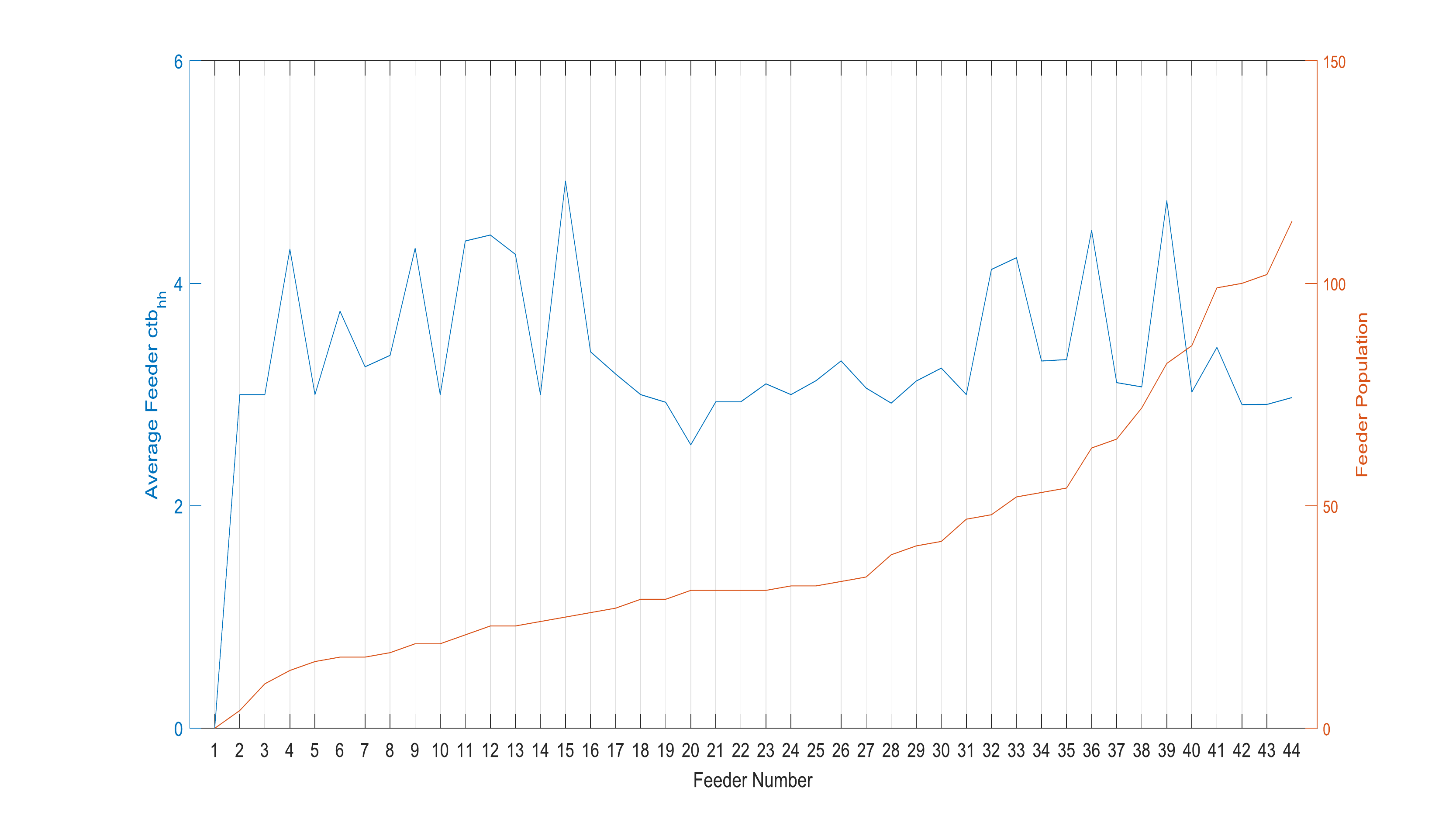}\\
\includegraphics[width=6.0in,height=2.5in]{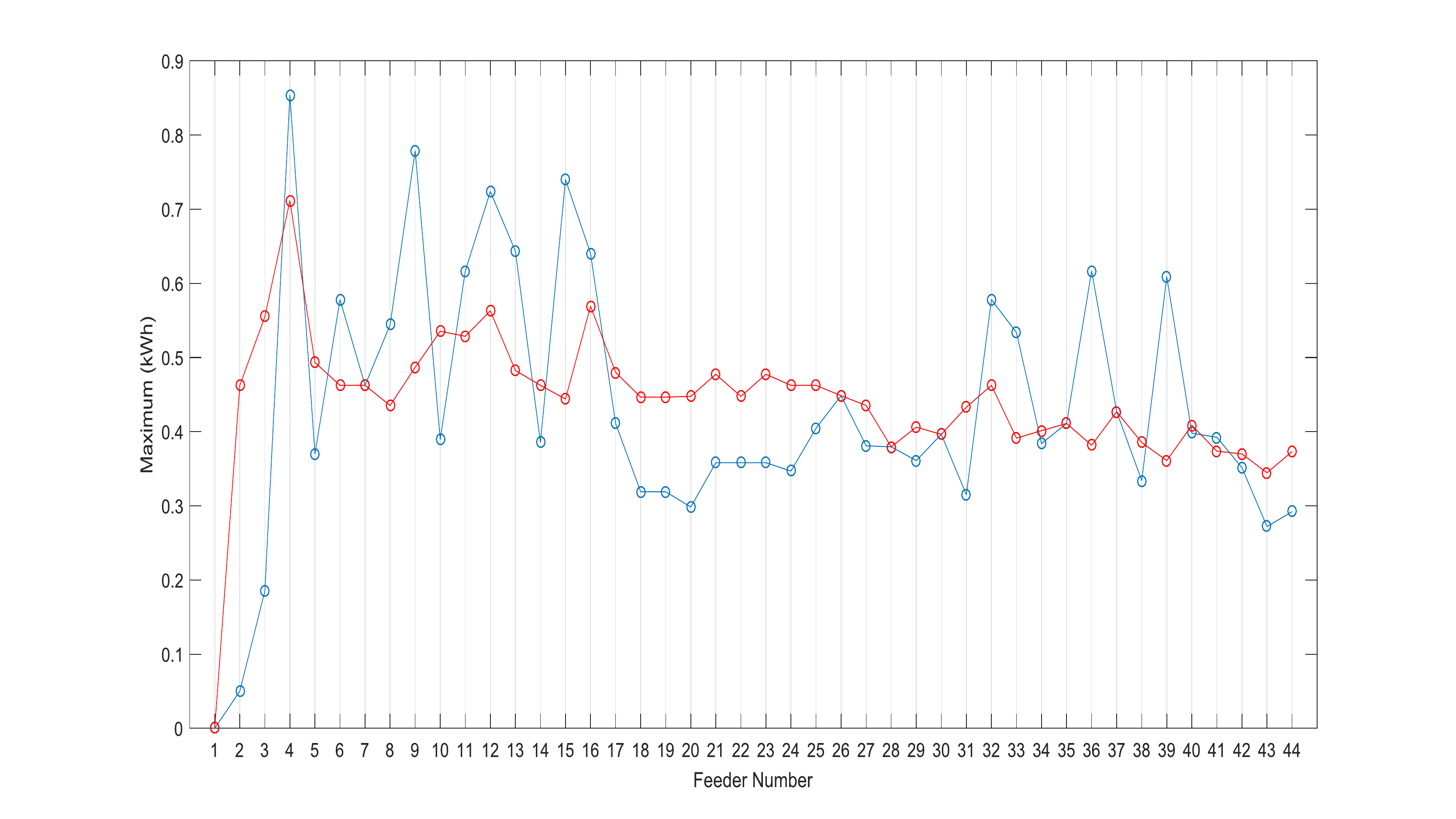}\\
\includegraphics[width=6.0in,height=2.5in]{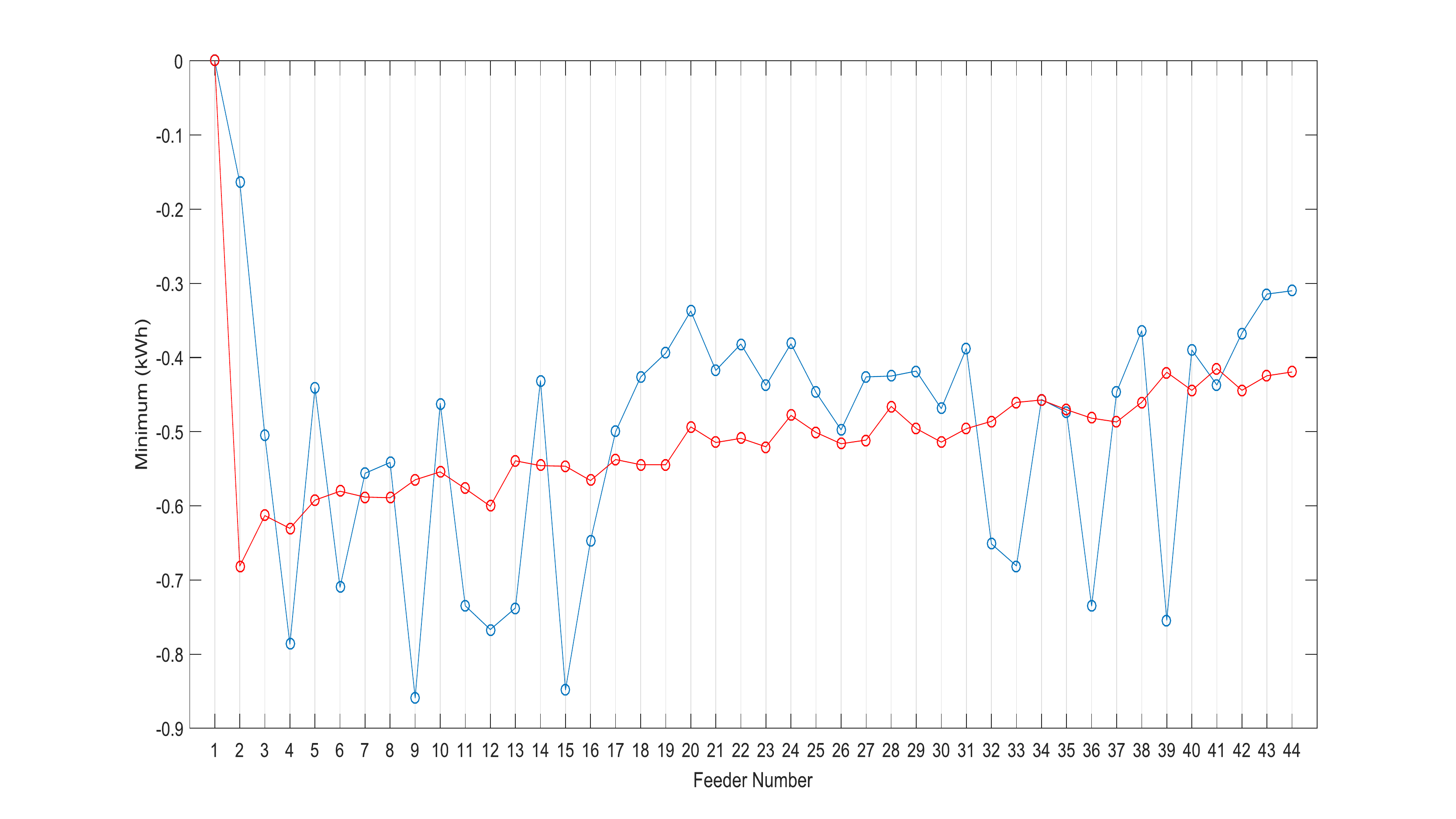}
\caption{Overview of results for all $44$ feeders. Top: The average $ctb_{hh}$ at each feeder (blue) and the variation in the feeder household population- the number of households along the feeder (red). Middle: The peak value of the feeder $90\%$ quantile, divided by the number of households along this feeder (study \textbf{a}). Bottom: The minimum value of the feeder $10\%$ quantile, divided by the number of households along this feeder (study \textbf{b}). Middle and bottom, red: Simulations randomly distributed seeds. Middle and bottom, blue: Simulations used CTBs to inform seed distribution. Note that feeder $1$ has only commercial properties.
\label{fallEVPV}}
\end{center}
\end{figure}

The confidence bounds results for the studies:
\begin{description}
  \item[a] $30\%$ EV uptake winter with
\begin{description}
  \item[i] seeds randomly distributed,
  \item[ii] seed distribution guided by CTB information,
\end{description}
  \item[b] $30\%$ EV and $30\%$ PV uptake summer with
\begin{description}
  \item[i] seeds randomly distributed,
  \item[ii] seed distribution guided by CTB information,
\end{description}
\end{description}
are summarised in Figure \ref{fallEVPV} for all $44$ feeders. Note that feeder $1$ is comprised of only commercial properties, so this feeder does not receive any LCTs. The first panel depicts each feeder's average $ctb_{hh}$ (blue) and household population (red). The middle image relates to study $a$, where the maximum value of the $90\%$ EV feeder quantile (the baseload subtracted), divided by the number of households along the feeder, is shown. The red and blue curve correspond to $a(i)$ and $a(ii)$ respectively. The bottom panel displays the results for study $b$, with the minimum value of the $10\%$ EV/PV feeder quantile (the baseload subtracted), divided by the number of households along the feeder, given. The red and blue trend are linked with $b(i)$ and $b(ii)$ respectively. There is an evident correlation between the feeder population and the red curves associated with $a(i)$ and $b(i)$. This behaviour was discussed in Section $\ref{cbs}$, where less populated feeders received larger LCT loads. When the CTB data is introduced, certain feeders attain greater extreme values, whilst at other feeders the load magnitude is reduced. This is demonstrated by the blue curve along the bottom two panels, which overall follows the top panel blue trend. This is expected since when CTBs are applied, the clustering algorithm favours feeders that have a higher proportion of larger properties. The feeders that receive amplified minimum and maximum values are especially vulnerable and further analysis at these sites is needed, such as determining the available headroom.

\section*{Acknowledgment} {This work was carried out with support of Scottish and Southern
Energy Power Distribution through the New Thames Valley Vision
Project (SSET203 New Thames Valley Vision) \url{www.thamesvalleyvision.so.uk}  funded by the Low
Carbon Network Fund established by Ofgem.}
\section*{References}
\bibliography{references}
\clearpage

\end{document}